\documentstyle[twoside,fleqn,espcrc2,psfig,epsfig]{article}
\def \gsim {\mbox{${}^> \hspace*{-7pt} _\sim$}}
\title{Four--neutrino Oscillations at SNO}

\author{C. Pe\~na-Garay\address{Inst. de F\'{\i}sica Corpuscular, 
C.S.I.C. - Univ. de Val\`encia, Spain}\thanks
{Talk presented at EURESCO Conference on Frontiers in Particle
 Astrophysics and Cosmology, San Feliu de Guixols, Spain. Based on 
M.~C.~Gonzalez-Garcia and C.~Pe\~na-Garay, hep-ph/0011245.}}
\begin{document}
\begin{abstract}
We discuss the potential of the Sudbury Neutrino Observatory (SNO) to 
constraint the four--neutrino mixing schemes favoured by the results 
of all neutrino oscillations experiments. 
Our results show that some information on the 
value of $\cos^2(\vartheta_{23}) \cos^2(\vartheta_{24})$ 
can be obtained by the first SNO measurement of the CC  
ratio, while considerable improvement on the knowledge of this mixing 
will be achievable after the measurement of the NC/CC ratio. 
\end{abstract}
\maketitle
\section{Introduction} 
The Sudbury Neutrino Observatory is a second generation
water Cerenkov detector using 1000 tonnes of heavy water, D$_2$O, 
as detection medium. SNO was designed to address the problem of the
deficit of solar neutrinos by having sensitivity to all flavours 
of neutrinos and not just to 
$\nu_e$, allowing for a model independent test of the oscillation 
explanation of the observed deficit. Such sensitivity can be achievable 
because energetic neutrinos can interact
in the D$_2$O of SNO via three different reactions. Electron neutrinos 
may interact via the Charged Current (CC) reaction. All 
non-sterile neutrinos may also interact via Neutral Current (NC) 
and via Elastic Scattering (ES) (smaller cross section).

The main objective of SNO is to measure the ratio of NC/CC events. 
In its first year of operation SNO is concentrating on the measurement
of the CC reaction rate while in a following phase, after the addition
of MgCl$_2$ salt to enhance the NC signal, it will also perform a precise
measurement of the NC rate. 
It is clear that a cross-section-normalized and acceptance-corrected ratio
higher than 1 would strongly indicate the oscillation of $\nu_e$ 
into $\nu_\mu$ and/or $\nu_\tau$. On the other hand a deficit on both
CC and NC leading to a normalized NC/CC ratio $1$, can
only be made compatible with the oscillation hypothesis if  
$\nu_e$ oscillates in to a sterile neutrino. 

Most of the studies of the potential of SNO have been performed in the 
framework of oscillations between two neutrino states where $\nu_e$ oscillates
into either an active, $\nu_e\to\nu_a$,  or a sterile, $\nu_e\to\nu_s$, 
neutrino channel. On the other hand, 
once the possibility of a sterile neutrino is considered, 
these two scenarios are only limiting cases of the most general 
mixing structure \cite{DGKK-99,ourfour} which permits 
simultaneous $\nu_e\to\nu_s$ and $\nu_e\to\nu_a$ oscillations. 
We consider those four--neutrino schemes favoured by considering together
with the solar neutrino data, the results of the two additional evidences 
pointing out towards the existence of neutrino masses and mixing: 
the atmospheric neutrino data and the LSND results. 
We concentrate on two SNO measurements: the first expected result on the 
CC ratio and the expected to be most sensitive, the ratio of NC/CC.  
The measurement of other observables,
such as the recoil energy spectrum of the CC events and the zenith
angular dependence can provide important information to distinguish
between the different allowed regions for $\nu_e$--active oscillations 
but they are not expected to be very sensitive as discriminatory between
the active and sterile oscillations.

\section{Two--Neutrino Mixing: Predictions for SNO}
\label{two}

The results presented in this section have been obtained for the  
oscillation parameters in the presently allowed regions to 
the solar neutrino problem in the two--neutrino schemes. 
The results of the global fit to the data 
 ( Ref.~\cite{concha} ) include 
(i) the SK data after  1117 days of operation (total number of events 
and day and night energy spectra ), (ii) Gallex, GNO and 
SAGE data, (iii) the Homestake data. We used the 
90\% and 99\% CL  regions found from the global minimum. 
The fit includes the latest 
standard solar model fluxes, BP00 model~\cite{BP00}. For details on the 
statistical analysis applied to the different observable we refer 
to Ref.~\cite{zsno}. 

The total number of events in the CC reaction at SNO is calculated 
with the $\nu$d CC cross section computed from the corresponding 
differential cross sections~\cite{kubodera}
 folded with the finite energy resolution 
function of the detector and integrated over the electron recoil energy.
For definiteness,we adopted the most optimistic 
total energy threshold $E_{th}=5~MeV$. The total number of events in 
the NC reaction at SNO is obtained using the $\nu$d NC cross 
section from~\cite{kubodera}.

In order to cancel out all energy independent efficiencies and normalizations 
we will use the ratio 
$R^{th}_{CC} = \frac{N^{th}_{CC}}{N^{SSM}_{CC}}\equiv{\mbox{[CC]}}$
where $N^{SSM}_{CC}$ is the predicted number of events in the case of 
no oscillations. The equivalent expression for the NC ratio
$R^{th}_{NC} = \frac{N^{th}_{NC}}{N^{SSM}_{NC}}\equiv{\mbox{[NC]}}$  
Out of those ratios one can compute the double ratio 
$\frac{R^{th}_{NC}}{R^{th}_{CC}}\equiv$ [NC]/[CC] for which the largest 
sources of uncertainties cancel out ~\cite{bkssno}. As it was shown 
in Ref.~\cite{kubodera}, the ratio between the NC and 
CC reaction cross sections is extremely stable against 
any variations of the inputs of the calculations. The expected 
total uncertainties for the [CC] ratio and the  [NC]/[CC] ratio are 6.7 \% and 
3.6 \% respectively assuming 5000 CC events and 1219 NC events~\cite{bkssno}.
\begin{figure}[htbp]
\centerline{\protect\hbox{\psfig{file=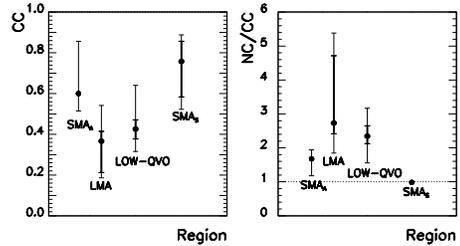,width=0.4\textwidth}}}
\vskip -1cm
\caption{[CC] and [NC]/[CC] predictions at SNO for the allowed regions 
in the two--neutrino mixing  scenarios obtained from the global analysis 
of solar neutrino data at 90 \% and 99 \%CL.}  
\label{2fsno}
\end{figure}

In Fig.~\ref{2fsno} we show the predicted [CC] and  [NC]/[CC] ratios 
for the allowed regions in the two flavour analysis. The dots 
correspond to the local best fit points and the error bars show the
range  of predictions for the points inside the 90 and 99 \%CL allowed regions.
The mapping of the regions onto these bars can be easily understood from 
the behaviour of the probability for the different solutions: \\
(a) For oscillations into active neutrinos the [NC]/[CC] ratio is 
simply the inverse of the [CC] prediction. Therefore: 
in the SMA region smaller mixing angles are mapped onto  
higher (lower) values of [CC]  ([NC]/[CC]) ratio. One may notice that 
the prediction for the 
[CC] rate for the global best fit point (0.60) is larger than the measured 
rate at Super--Kamiokande. This is due to the nearly flat spectrum 
at Super--Kamiokande which implies that the best fit point in the global 
analysis corresponds to a smaller mixing angle than the best fit point 
for the analysis of rates only; in the LMA region, 
the lower $\Delta m^2$ and $\theta$ values 
are mapped onto higher (lower) [NC]/[CC]  ([CC]) ratios and viceversa; 
in the LOW region the higher (lower) [NC]/[CC]  ([CC]) ratio occurs 
for smaller $\theta$ and higher $\Delta m^2$. 

(b) For the sterile case, the best fit point in SMA occurs at 
lower $\Delta m^2$ than in the active case and this produces a higher 
prediction for the [CC] ratio (0.76). The [NC]/[CC] ratio 
takes an almost constant value very close to one (0.98 in the best fit point),
since both numerator and denominator are proportional to 
$\langle P_{\nu_e\to\nu_e}\rangle$. 
It is smaller than one because for the SMA solution the probability increases 
with energy in the range of detection at SNO and the threshold for the 
NC reaction is below the one for the CC one.

What we see from these results is that while the data on [CC]  can give 
a hint towards large or small mixing solutions, it  will be hard to 
distinguish active from sterile oscillations on the only bases of this
measurement. This is not the case for the [NC]/[CC] ratio where 
both scenarios appear nicely separated. It is not hard to foresee 
from these results that from the [NC]/[CC] measurement SNO will be able to 
constraint the additional mixings in the four--neutrino scenario which 
describe the admixture of active and sterile oscillations. This is the 
main point in this paper.

\section{Allowed Four--neutrino Mixing Parameters}
\label{fourana}
Together with the results from the solar neutrino  
experiments we have two more evidences pointing out towards the existence of  
neutrino masses and mixing: the atmospheric neutrino data 
and the LSND results. All these experimental results can be 
accommodated in a single neutrino oscillation framework only if there 
are at least three different scales of neutrino mass-squared differences. 
The simplest case of three independent mass-squared differences 
requires the existence of a light sterile neutrino, 
{\it i.e.} one whose interaction with 
standard model particles is much weaker 
than the SM weak interaction, 
so it does not affect the invisible Z decay  
width, precisely measured at LEP.

There are six possible four--neutrino schemes that can accomodate all these
evidences. They can be divided in two classes: 3+1 and 2+2. In the 
3+1 schemes there is a group of three neutrino masses separated from an 
isolated mass by a gap of the order of 1eV which gives the mass-squared 
difference responsible for the short-baseline oscillations observed in the 
LSND experiment. In 2+2 schemes there are two pairs of close masses separated 
by the LSND gap. 3+1 schemes are disfavoured by experimental data 
with respect to the 2+2 schemes but they are still marginally allowed.

As discussed in Ref.~\cite{zsno}, for 
any of these four--neutrino schemes, either 2+2 or
3+1, only four mixing angles are relevant in the study of solar neutrino 
oscillations~\cite{DGKK-99,ourfour,GL}.
The survival of solar $\nu_e$'s 
mainly depends on the mixing angle 
$\vartheta_{12}$, 
whereas the mixing angles 
$\vartheta_{23}$ and $\vartheta_{24}$ 
determine the relative amount of transitions into sterile $\nu_s$ 
or active $\nu_a$, this last one being a combination of 
$\nu_\mu$ and $\nu_\tau$ controlled by the mixing angle $\theta_{34}$. 
$\nu_\mu$ and $\nu_\tau$ 
cannot be distinguished in solar neutrino experiments, 
because their matter potential 
and their interaction in the detectors are equal, 
due only to NC weak interactions. 
As a consequence the active/sterile ratio  and the survival 
probability for solar 
neutrino oscillations do not depend on the mixing angle 
$\vartheta_{34}$, and depend on the mixing angles 
$\vartheta_{23}$ 
$\vartheta_{24}$ only through the combination 
$\cos{\vartheta_{23}} \cos{\vartheta_{24}}$. For
further details see Ref.~\cite{DGKK-99,ourfour}.

In the general case of simultaneous $\nu_e\to\nu_s$ and $\nu_e\to\nu_a$ 
oscillations 
the corresponding probabilities are given by~\cite{DGKK-99,ourfour}
\begin{eqnarray} 
&& 
P_{\nu_e\to\nu_s} 
= 
c^2_{23} c^2_{24} 
\left( 1 - P_{\nu_e\to\nu_e} \right) 
\,, 
\label{Pes} 
\label{Pea} 
\\ 
&& 
P_{\nu_e\to\nu_a} 
= 
\left( 1 - c^2_{23} c^2_{24} \right) 
\left( 1 - P_{\nu_e\to\nu_e} \right) 
\,. 
\end{eqnarray} 
where  $P_{\nu_e\to\nu_e}$ takes the standard two--neutrino oscillation
form for $\Delta m^2_{12}$ and $\theta_{12}$ but
computed with the modified matter potential 
$A \equiv A_{CC} + c^2_{23}c^2_{24} A_{NC}$.
Thus the analysis of the solar neutrino data in the
four--neutrino mixing schemes is equivalent to the two--neutrino
analysis but taking into account that the parameter space is now 
three--dimensional $(\Delta m^2_{12},\tan^2\vartheta_{12}, 
\cos^2{\vartheta_{23}} \, \cos^2{\vartheta_{24}})$. 
We want to stress that, although originally this derivation 
was performed in the framework of the 2+2 schemes~\cite{DGKK-99,ourfour},
it is equally valid for the 3+1 ones \cite{GL}.

The allowed regions in the three--parameter space for the global 
combination of observables are shown in Ref.~\cite{concha}. 
The global minimum used in the construction of the 
regions lays in the LMA region and for pure $\nu_e$--active oscillations, 
$c_{23}^2c_{24}^2=0$. The SMA region is always a valid solution  
for any value of $c_{23}^2c_{24}^2$ at 95\% CL 
( 0.11$<c_{23}^2c_{24}^2<$0.31 allowed at 90\% CL). 
On the other hand, the LMA and LOW--QVO solutions disappear for 
increasing values of the mixing $c_{23}^2c_{24}^2$ (0.72 and 0.76 at 99\% CL
, respectively).

\section{Expected Rates at SNO in Four--Neutrino Schemes}
\label{foursno}
In this section, we present the predictions for the CC ratio and for 
the NC/CC ratio in the four--neutrino scenario previously described. 
This scenario contains as limiting cases the pure $\nu_e$--active 
and $\nu_e$--sterile neutrino oscillations. 

\begin{figure}[htbp]
\centerline{\protect\hbox{\psfig{file=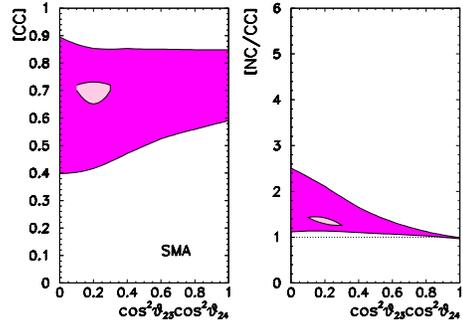,width=0.4\textwidth}}}
\vskip -1cm
\caption{[CC] and [NC]/[CC] predictions at SNO as for the SMA region 
in the four--neutrino scenario obtained from the global analysis of 
solar neutrino data at 90 \% (lighter) and 99 \%CL (darker). The dotted 
line corresponds to the prediction in the case of no oscillations.}  
\label{4fsno_sma}
\end{figure}

\begin{figure}[htbp]
\centerline{\protect\hbox{\psfig{file=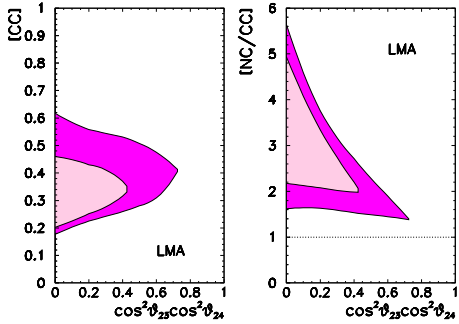,width=0.4\textwidth}}}
\vskip -1cm
\caption{Same as Fig.~\protect{\ref{4fsno_sma}} for the LMA region.}
\label{4fsno_lma}
\end{figure}

\begin{figure}[htbp]
\centerline{\protect\hbox{\psfig{file=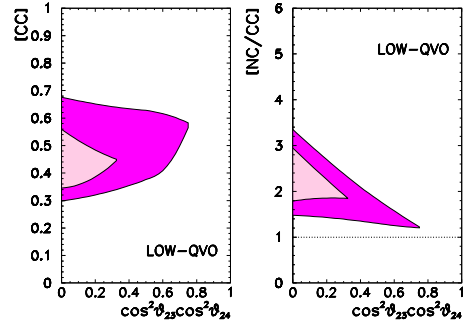,width=0.4\textwidth}}}
\vskip -1cm
\caption{Same as Fig.~\protect{\ref{4fsno_sma}} for the LOW region.}
\label{4fsno_low}
\end{figure}

In Figs.~\ref{4fsno_sma}$-$\ref{4fsno_low} we show the results for the 
predicted [CC] ratio and [NC]/[CC] ratio for the different allowed 
regions (SMA, LMA, LOW-QVO) at 90 and 99 \%CL as a function 
of $c_{23}^2c_{24}^2$. 
The general behaviour of the dependence of the predicted ratios with 
$c_{23}^2c_{24}^2$ can be easily understood using the following 
simplified expressions:
\begin{eqnarray}
\mbox{\rm [CC]} &\sim& 
 P_{\nu_e\to\nu_e} \; , \label{CC}\\
\frac{\mbox{\rm [NC]}}
{\mbox{\rm [CC]}} &\sim &
\frac{1 - c^2_{23} c^2_{24}(1 - P_{\nu_e\to\nu_e})}{P_{\nu_e\to\nu_e}}\;. 
\label{NCCC}
\end{eqnarray}
From Eq.~(\ref{CC}), the only dependence of [CC] on 
$c_{23}^2c_{24}^2$  is 
due to the modification of the matter potential entering in  
the evolution equation and it is very weak. 
The dependence of the allowed range of the [CC] ratio with 
$c_{23}^2c_{24}^2$ displayed in the figures
arises mainly from the variation of the size of the allowed regions. 
Alternatively following Eq.~(\ref{NCCC}) we find a stronger linear 
dependence of [NC]/[CC] on $c_{23}^2c_{24}^2$ with slope  
$(- 1 + P_{\nu_e\to\nu_e})/P_{\nu_e\to\nu_e}\sim 1-1/$[CC] 
and intercept $1/P_{\nu_e\to\nu_e}\sim 1/$[CC].
This simple description is able to reproduce the main features of our 
numerical calculations as can be seen in the figures. 

Figs.~\ref{4fsno_sma}$-$\ref{4fsno_low} contain the main quantitative
result of our analysis in the four--neutrino mixing scenario. From each 
of them it is possible to infer the allowed range of the active--sterile 
admixture, $c_{23}^2c_{24}^2$, compatible, within the expected uncertainty, 
with a given SNO measurement of the ratios. 
Also, comparing the allowed ranges for the different solutions 
one can study the potential of these measurements as discriminatory 
among the three presently allowed regions. Of course, both issues are 
not independent as we have no ``a priory'' knowledge of which is the
right solution and both must be discussed simultaneously. 
In order to do so we pass to describe and compare in detail the 
predictions in the different regions.

The results for the SMA solution are shown in Fig.~\ref{4fsno_sma}.a 
and~\ref{4fsno_sma}.b for [CC] and  [NC]/[CC] ratios respectively. 
First we notice that we find a small region allowed at 90 \%CL only for 
a non--vanishing admixture of active and sterile oscillations as mentioned 
before. In this region [CC]$\sim$0.65--0.73 and 
[NC]/[CC]$\sim$ 1.3--1.4. The predictions at 99\% range from 
[CC]$\sim$ 0.4--0.9 ([NC]/[CC]$\sim$1.1--2.5) 
for pure $\nu_e$--active scenario to 
to [CC]$\sim$ 0.59--0.85 ([NC]/[CC]$\sim$0.96--0.98) for pure 
$\nu_e$--sterile oscillations.
Thus if SNO observes a ratio [CC]$<$0.58 the 
value of $c_{23}^2c_{24}^2$ can be constrained to be smaller than 1 
disfavouring pure $\nu_e$--sterile oscillations. On the contrary 
a measurement of [CC]$\gsim$0.68 will immediately hint towards 
the SMA solution but will not provide any information on the active--sterile
admixture. Also, one must notice, that such value, although allowed
by the present global statistical analysis at 99 \%CL, will imply
a strong disagreement with the total rate event rate observed at 
Super--Kamiokande.

As seen in Fig.~\ref{4fsno_sma}.b the [NC]/[CC] ratio is more sensitive 
to the active--sterile admixture. 
To guide the eye, in the figures for the 
[NC]/[CC] ratio we plot a dotted line for the prediction in the case of 
no oscillation [NC]/[CC]=1. For any of the solutions,  
the allowed range for this ratio shows as general behaviour a decreasing 
with $c_{23}^2c_{24}^2$ due to two effects: $(i)$ the allowed regions become
smaller and $(ii)$ the prediction decreases when more sterile neutrino 
is involved in the oscillations as described in Eq.~(\ref{NCCC}). The 
measurement of higher values of this ratio will favour the four--neutrino
scenario with larger component of $\nu_e$--active oscillations. On the
other hand a measurement of  [NC]/[CC]$\sim 1$, will push the 
oscillation hypothesis towards the pure 
$\nu_e$--sterile oscillation scenario. This case will 
be harder to differentiate from the non--oscillation scenario. 
We find that with the expected sensitivity the parameter 
$c_{23}^2c_{24}^2$ is constrained to be above 0.44 at 99\% CL and that 
the pure $\nu_e$--active oscillations in the SMA region are compatible 
with [NC]/[CC]= 1 only at $\sim 5\sigma$.
The predictions for oscillation parameters in the LMA region are
shown in Figs.~\ref{4fsno_lma}.a and~\ref{4fsno_lma}.b for 
[CC] and [NC]/[CC] ratios respectively. 
The predictions at 99\% vary in the range 
[CC]$\sim$ 0.18--0.62  and [NC]/[CC]$\sim$ 1.4--5.6. 
The first thing we notice by comparing Fig.~\ref{4fsno_lma}.a
with Fig.~\ref{4fsno_sma}.a and Fig.~\ref{4fsno_low}.a 
is that the most discriminatory scenario for the [CC] rate results if 
SNO finds a small value [CC]$\sim 0.25$. This would 
significantly hint towards the LMA solution to the solar 
neutrino problem and towards and $\nu_e$--active oscillation scenario. 
First, it is well separated from the predictions for the SMA and LOW 
regions. Second, it will include as a bonus a small but measurable 
day--night asymmetry. And 
third it will constrain the $c_{23}^2c_{24}^2$ to a small value 
($\sim 0.2$). On the contrary the less discriminatory scenario will be a 
measurement 0.4$<$[CC]$<$0.6 where the prediction would 
be compatible with both SMA and LOW-QVO solutions and no improvement 
on our knowledge of the four--neutrino schemes is possible.
The [NC]/[CC] ratio can definitively improve the discrimination between
the different scenarios provided its measurement lays in the upper range. 
For instance a measurement of 
[NC]/[CC]$\sim$ 4 ($\pm$ 0.7 at $5\sigma$) will be 
conclusive for selecting LMA as the solution to the SNP and will imply
and upper bound on  $c_{23}^2c_{24}^2<0.3$. 
The predictions for the LOW--QVO region lay between the ones for SMA and LMA
as displayed in Fig.~\ref{4fsno_low} and therefore they are more difficult
to discriminate.  The predictions at 99\% vary in 
the range [CC]$\sim$ 0.3--0.68  and [NC]/[CC]$\sim$ 1.2--3.4. 
As a consequence we see that a low [CC] ratio but still within the 
99\% CL range allowed for this region, 0.3$<$[CC]$<$0.4, 
will constrain significantly 
the $c_{23}^2c_{24}^2$ parameter compatible with this solution but 
it  will not be distinguishable from the LMA solution unless the measured 
[CC]$<$0.3. As mentioned above, 
the [NC]/[CC] ratio will be able to differentiate the LMA and LOW-QVO 
solutions if not in the range [1.5,3]. One should also notice that for the 
upper part of this range a positive measurement of the day--night asymmetry 
and the zenith dependence will point towards the higher 
$\Delta{m}^2$ of the LOW region as the solution.

This work was supported by grants DGICYT-PB98-0693 and PB97-1261, 
GV99-3-1-01, EU network ERBFMRXCT960090 and ESF network 86.


\end{document}